\documentclass[11pt,a4paper]{article}
\usepackage[utf8]{inputenc}
\def\parfrac#1#2{{\left(\frac{#1}{#2}\right) }} 
\usepackage{caption}
\usepackage{jcappub}
\usepackage{bm}
\usepackage{epsfig}
\usepackage{bbold}
\usepackage{graphicx,subfig}
\usepackage{amsmath}
\usepackage{amssymb}
\usepackage{amsbsy}
\usepackage{color}
\usepackage{slashed}
\usepackage{afterpage}
\usepackage{psfrag}
\usepackage[noabbrev]{cleveref}
\usepackage{dsfont}
\usepackage{enumerate}
\usepackage[shortlabels]{enumitem}
\usepackage[utf8]{inputenc}
\usepackage{amsmath}
\usepackage[dvipsnames]{xcolor}
\usepackage{graphicx}
\usepackage{braket}
\usepackage{float}

\title{\begin{center}
Stupendously Large \\
Primordial Black Holes from \\ the QCD axion
   \end{center}}

\author[a]{Miguel Faria,}
\author[a]{Ricardo Z. Ferreira}
\author{and}
\author[b,c]{Fabrizio Rompineve}

\affiliation[a]{CFisUC, Department of Physics, University of Coimbra, Rua Larga P-3004-516, Coimbra, Portugal}
\affiliation[b]{Departament de F\'isica, Universitat Aut\`onoma de Barcelona, 08193 Bellaterra, Barcelona, Spain
\looseness=-1}
\affiliation[c]{Institut de F\'isica d’Altes Energies (IFAE) and The Barcelona Institute of Science and Technology (BIST), Campus UAB, 08193 Bellaterra (Barcelona), Spain}%\looseness=-1}

\emailAdd{miguelfaria@uc.pt}
\emailAdd{rzferreira@uc.pt}
\emailAdd{frompineve@ifae.es}

\makeatletter
\gdef\@fpheader{}
\makeatother

\abstract{The inflationary diffusion of (pseudo-)scalar fields with discrete symmetries can seed the formation of a gas of closed domain walls after inflation, when the distance between degenerate minima in field space is not too far from the inflationary Hubble scale. Primordial black holes (PBHs) can then be formed once sufficiently heavy domain walls re-enter the Hubble sphere. In this scenario, inflation determines a distinctive PBH mass distribution that is rather flat and can thus lead to a sizable total abundance of PBHs, while avoiding some of the downsides of PBH formation from critical collapse.
We show that generic QCD axion models, with decay constant close to the inflationary Hubble scale, can yield up to $1\%$ of the dark matter (DM) today in the form of PBHs, while being compatible with isocurvature constraints from Cosmic Microwave Background observations. This occurs for values of axion decay constants around $f_a\simeq 10^{8}~\text{GeV}$, that is the region targeted by axion helioscopes and partially constrained by astrophysical observations. The resulting PBHs have \textit{stupendously} 
large masses, above $10^{11}M_\odot$, and their existence can be probed by Large Scale Structure observations. Larger PBH abundances can be generated by axion-like particles.
Alternatively, in scenarios where isocurvature constraints can be relaxed, we find that the totality of the DM can be produced by the QCD axion misalignment mechanism, accompanied by a ${\cal O}(10^{-3})$ DM fraction in PBHs of masses $(10^5-10^6)~M_\odot$. These can act as seeds for the formation of massive black holes at large redshifts, as suggested by recent JWST observations. 
}

\begin{document}

\maketitle

\section{Introduction}

Inflation is known to diffuse light scalar fields in a stochastic process famously described by Starobinsky \cite{Starobinsky:1986fx}. In each e-fold of inflation,
the field is diffused by an amount $H/(2\pi)$  in a random walk that leads to a Gaussian probability distribution with a variance that increases with the square root of the number of efolds \cite{Starobinsky:1986fx,Linde:1990yj}.
In certain cases, the diffusion can occasionally take the field far away from its starting point at the beginning of the observable period of inflation. When the field enjoys a discrete symmetry, these newly populated regions of field space can eventually correspond to a different, degenerate minimum  \cite{Linde:1994wt,Hasegawa:2018yuy,Kitajima:2020kig}, as shown in Fig.~\ref{fig:Sketch of the mechanism}. Closed domain walls separating these regions then necessarily form, once the field becomes massive, and collapse once they reenter the Hubble horizon after inflation.
Due to its inflationary origin, such a domain wall gas is long-lived even in axion (and axion-like) scenarios with domain wall number equal to unity (such as in the simplest KSVZ scenario \cite{Kim:1979if,Shifman:1979if}), in contrast to the commonly studied post-inflationary axion string-wall network, see e.g.~\cite{Kawasaki:2014sqa}. This makes the phenomenon roughly independent of the specific model, as long the distance between the minima is not far from the diffusion jump. 

In this work, we revisit this mechanism of domain wall formation primarily focusing on the QCD axion, and secondly on axion-like-particles (ALPs).~\footnote{In axion models, cosmic strings might also be formed if the diffusion is large enough \cite{Lyth:1991ub,Redi:2022llj,Gorghetto:2023vqu} but here we consider the limit where the symmetry is broken during inflation, $H_I/(2\pi) \ll f$, such that strings have been sufficiently diluted once the observable scales exited the horizon during inflation.} As we shall see, we will uncover interesting observational implications that have not yet been explored. 
In particular, following on previous related works \cite{Hasegawa:2018yuy,Kitajima:2020kig}, we compute for the first time the dark matter abundance that results from the post-inflationary collapse of such a gas of domain walls.~\footnote{The possibility of forming PBHs by the same mechanism has previously been considered for ALPs by~\cite{Sakharov:2021dim}, in relation to the stochastic GW signal at Pulsar Timing Arrays (PTAs). Nonetheless, the authors have assumed spherical symmetry throughout their work, as well as neglected the abundance of dark matter produced by the decay products of the walls that do not collapse to PBHs.}
The dark matter relics from the walls have two components: first, there are axion quanta coming from the collapse of the DWs; second, there can be Primordial Black Holes (PBHs) that originate from Hubble-sized DWs that have Schwarzschild radius larger than or equal to the Hubble radius.  
Interestingly, as we will show in more detail, whenever the total dark matter abundance is substantial, the majority of it turns out to be in the form of PBHs. Furthermore, we will show that the PBH mass spectrum is predicted to be rather flat, in sharp contrast with most other models of PBH formation, both from critical collapse and from the collapse of DW networks that form from spontaneous symmetry breaking in the post-inflationary Universe~\cite{Vachaspati:2017hjw,Ferrer:2018uiu,Ferreira:2024eru}. On the other hand, this feature is similarly realized in scenarios where domain walls are nucleated during inflation via quantum tunneling~\cite{Garriga:1992nm}.

Besides the observational constraints on the PBH fraction, we also study the isocurvature constraints on the scenario. These are known to be intrinsic to the model \cite{Linde:1990yj}, and here we provide an updated analysis that is valid beyond the $\theta \ll 1$ limit. However, as we will argue, these constraints can be alleviated with the addition of a simple layer of model building.

This work is organized as follows. In Section \ref{sec: Dark matter and Primordial Black Holes} we present the mechanism and provide the basic formulae to calculate the abundance of dark matter and primordial black holes from the collapse of the domain walls. Then, in Section \ref{sec: Isocurvatures} we compute the amount of isocurvatures perturbations at CMB scales. In Section \ref{sec: Results} we specify the results to the case of the QCD axion, then extend them to a general ALP. Finally, in Section \ref{sec: Evading Isocurvatures} we discuss how the observational constraints on isocurvatures can be evaded and in Section \ref{sec: Conclusions} we conclude.

\section{Dark matter from inflationary domain walls}
\label{sec: Dark matter and Primordial Black Holes}

\subsection{Domain walls from inflationary diffusion}
Inflation is known to efficiently dilute topological defects. For example,  cosmic strings that form when the $U(1)$ Peccei-Quinn symmetry \cite{Peccei:1977hh} is broken during inflation, i.e. when $H_I \ll 2 \pi f_a$ where $H_I$ is the Hubble scale during inflation and $f_a$ is the axion decay constant, are diluted by the inflationary expansion.~\footnote{We consider the PQ symmetry to be broken during inflation if the radial mode is sufficiently heavy. In a quartic potential the radial mode has a mass $m_r = \sqrt{2 \lambda} v$, where $\lambda$ is the quartic coupling and $v$ the VEV, related to the axion decay constant by $v= f_a N_{dw}$ with $N_{dw}$ the domain wall number. Therefore, we work under the assumption that $H_I \ll \sqrt{2 \lambda} f_a N_{dw}$.} Yet, the massless Goldstone boson associated with the angular direction, the axion, can still yield non-trivial dynamics as it undergoes a random walk impinged by de-Sitter quantum fluctuations of size  $ H_I/(2\pi)$ per e-fold. In general, the same dynamic also holds for a more general ALP as long as its mass $m_a$ is light enough during inflation $m_{a} \ll H_I$.

\begin{figure}
    \centering
\includegraphics[width=\textwidth]{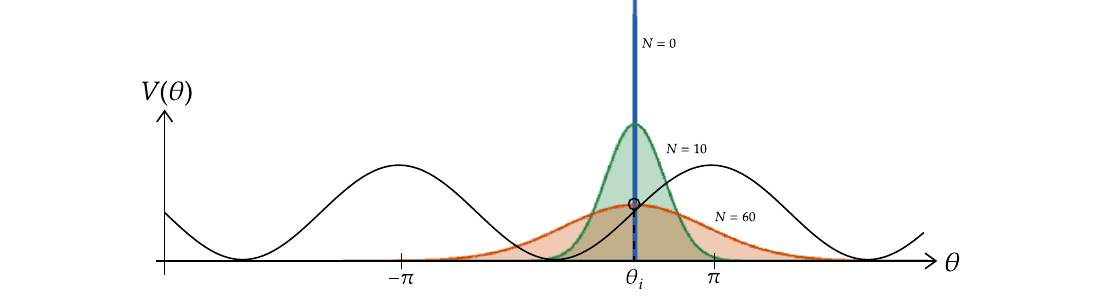}
    \caption{Sketch representation of the diffusion of the axion throughout inflation  according to the probability distribution $p(N,\theta)$ in Eq. \ref{eq: Diffusion probability}. 
    Different colors represent different time instances (e-fold number). The black solid line denotes instead the would-be axion cosine potential after inflation.}
    \label{fig:Sketch of the mechanism}
\end{figure}

The effect of diffusion is to cumulatively displace the axion field $a$ away from its initial value $a_i$ at the beginning of the observable period of inflation, according to the probability \cite{Linde:1990yj}
 \begin{eqnarray} \label{eq: Diffusion probability}
 	p(N,\theta) = \frac{1}{\sqrt{2\pi}\sigma_{\theta}(N)}  \exp\left(-\frac{(\theta-\theta_i)^2}{2\sigma_{\theta}^2(N)}\right)
 \end{eqnarray}
 where $\theta \equiv a/f$ is the angle and 
 \begin{eqnarray}
 \sigma_{\theta}(N)=\frac{y}{2\pi }   \sqrt{N} \label{eq: diffusion variance}
 \end{eqnarray}
 with $y=H_I/f$, is the variance caused by the diffusion throughout $N$ e-folds (steps). We give a pictorial description of the mechanism in Fig. \ref{fig:Sketch of the mechanism}.

Smaller scales undergo more diffusion because they exit the horizon at a later time, i.e. at larger $N$, during inflation and so can explore a larger field range. 
In the particular case of axions, if the diffusion is large enough, there can be regions of space where the field explores field values $|a| > \pi f$,
i.e. beyond the maximum of the (would be) axion potential. These are the regions that will later be enclosed by DWs.

The total volume of space $V_{\theta>\pi}(N)$ that reached $\theta >\pi$ is given by
 \begin{eqnarray}
 	V_{|\theta|>\pi}(N) = V(N)  f(N), \quad   f(N) \equiv \int_{|\theta| > \pi} 	p(N,\theta) \, d \theta = 1 - \text{Erf}(\Theta)
 \end{eqnarray} 
where $f(N)$ the fraction of total volume, $\text{Erf}(x)$ is the error function and $\Theta \equiv (\pi-|\theta_i|)/(\sqrt{2} \sigma_{\theta}(N))$.  This volume grows at a rate \cite{Hasegawa:2018yuy,Kitajima:2020kig}
 \begin{eqnarray}
 	\frac{d V_{|\theta|>\pi}(N)}{dN} = 3 V_{|\theta|>\pi}(N) + V(N) \frac{df(N)}{dN} \,
 \end{eqnarray} 
 where the first term on the right-hand side corresponds to the growth of regions where $|\theta|$ was already larger than $\pi$, and the second term is the rate at which new regions with $|\theta|> \pi$ are created. 
 Therefore, the fraction of volume that corresponds to newly formed regions with $|\theta|>\pi$ at the e-fold $N$ is 
 \begin{eqnarray} 
 \label{eq: probability of forming walls}
 	\beta(N)= \frac{ V(N) \frac{df}{dN}}{V(N)} = \frac{1}{\sqrt{\pi}N} \Theta(N)  e^{-\Theta(N)^2} \,.
 \end{eqnarray}

 \subsection{Dark matter from inflationary domain walls}

After inflation, when the condition $m_a = H$ is satisfied, 
  the field starts oscillating around the minimum. In most regions it will oscillate around $a=0$, however, in the regions of space where the field has diffused to values    $|a|> \pi f_a$,  it will oscillate around a different minimum of the axion potential.\footnote{In axion models with $N_\text{dw}=1$, all the minima are actually physically equivalent to the minimum at 0, whereas if $N_\text{dw}>1$ the $N_\text{dw}$ vacua in one fundamental periodicity are physically distinct. Domain walls form in both cases.} Connecting these regions of different minima there will necessarily be domain walls of thickness $\delta \simeq 1/m_a$ and 
  whose distribution of sizes is controlled by $\beta(N)$ in Eq. \ref{eq: probability of forming walls}. The DWs are closed and enclose the region with $|\theta|> \pi$. They can even be nested if the field excursion is large enough such that $\theta >3 \pi$, however, in our scenario such situations are exponentially suppressed and thus very unlikely. We also stress that these  DWs are not bounded by axion strings, which are absent in this scenario as long as the radial mode is sufficiently heavy during inflation. 
  Domain Walls of this type are somewhat analogous to those that arise in the post-inflationary QCD axion scenario after the decay of the string network propels the axion field to larger values~\cite{Gorghetto:2020qws}.

Once the domain walls are formed there are two different dynamical regimes. While the DWs have superhorizon sizes, they are essentially frozen/comoving with the expansion. However, once they enter the horizon they are expected to quickly collapse due to their own tension (or vacuum energy, in case there is a small bias between the minima). 
We shall assume that this collapse typically results in the production of non-relativistic (or mildly relativistic) axions (following the results of~\cite{Kawasaki:2014sqa})~\footnote{However, notice that the results of~\cite{Kawasaki:2014sqa} are based on simulations of an axion string-wall networks in the post-inflationary scenario. In the case of interest for this work, walls may achieve larger $\gamma$ factors than in~\cite{Kawasaki:2014sqa}, thereby possibly resulting in more relativistic axions upon collapse. We leave this interesting question for future work.}, unless the re-entering wall is heavy enough to form a primordial black hole (PBH), as we shall investigate at the end of this section. In both cases, the collapsing walls produce dark matter species.

To compute the abundance of dark matter from the collapsing DWs we need to account for
all the closed domain walls that reenter the Hubble sphere at different times. We model the axion potential as 
\begin{equation}
V(a)=m_a(T)^2 f_a^2\left[1-\cos\left(\frac{a}{f_a}\right)\right],
\end{equation}
where $m_a(T)$ is the temperature-dependent axion mass and $f_a$ the axion decay constant. Additionally, we consider the possibility of a small vacuum energy difference between the minima $\Delta V\ll m_a^2 f_a^2$ (also referred to as bias), that can be generated from an explicit symmetry-breaking term. The resulting DW tension is then $\sigma_{\text{dw}}=8 m_a(T) f_a^2$ (for the QCD axion this slightly underestimates the actual tension, but the difference is not relevant for our conclusions) \cite{GrillidiCortona:2015jxo}.
The computation is then divided into the following four steps. First, we estimate the energy density $\rho_{\text{dw}}$ of a DW that enters the horizon at the $t_*$ to be given by $\rho_\text{dw}=\sigma_{\text{dw}} \cdot \text{Area}_H/\text{Vol}_H + \Delta V $, where $\text{Area}_H$, $\text{Vol}_H$ are, respectively, the area and volume enclosed by the Hubble sized DW. For the purpose of estimating the total energy carried by a collapsing wall (and only to this aim, i.e. we will crucially not make this assumption when we estimate the PBH fraction), we model the DW as spherical objects of radius $1/H_*$ where the subscript $*$ denotes quantities evaluated at $t_*$. We allow for temperature dependence of the DW tension $\sigma_{\text{dw}}$, 
which is relevant for cases such as the QCD axion.

Second, we multiply the energy density by the probability that a domain wall of radius $1/H_*$ enters the horizon at the time $t_*$. This probability is controlled by $\beta(N_*)$ in Eq. \ref{eq: probability of forming walls} where $N_*$ is the e-fold number that is associated with the comoving scale $k_*$ that re-enters the horizon at $k=a_* H_*$. 

To relate the scale $k_*$ to $N_*$ we note that using entropy conservation
\begin{eqnarray}
    N_* = \log \left( \frac{a_* H_*}{a_0 H_0} \right) = \log\left( \frac{a_\text{eq} H_\text{eq}}{a_0 H_0} \right) + \log \left[ \left( \frac{g_{s,eq} }{g_{s,*}}\right)^{1/3} \left( \frac{g_{\rho,*} }{g_{\rho,eq}}\right)^{1/2} \frac{T_*}{T_\text{eq}} \right] \, .
\end{eqnarray}
where   ($g_s$) $g_\rho$ is the number of (entropic) relativistic degrees of freedom and the subscripts $0,$ and eq denote quantities evaluated today and at matter-radiation equality (respectively). Plugging the identity $a_{eq} H_{eq}/(a_0 H_0) = 219 \Omega_0 h$ \cite{Liddle_2003}, where $\Omega_0=0.315$ is the matter density parameter today \cite{Planck:2018vyg}, and the values of $g_{s,eq} = 3.94, g_{\rho,eq}=3.36$
one then finds
\begin{eqnarray} \label{eq: N_* (T_*)}
N_* = 3.68 + \frac{1}{2} \log{g_{\rho,*}} - \frac{1}{3}\log{g_{s,*}} + \log{\frac{T_*}{T_{eq}}} \, .
\end{eqnarray}

Third, we take the DW collapse process to occur on time scales that are much smaller than a Hubble time, and thus effectively redshift the energy density in the axion field as non-relativistic matter from the time when the walls re-enter the Hubble sphere until the present time. This approximation is justified by the fact that the collapsing domain wall quickly reaches ultrarelativistic speeds \cite{Ferreira:2024eru}, and remains valid even when DW collapse results into PBH formation, as we will discuss in the next section.

Fourth, we restrict the analysis to situations where the large majority of the DWs enter the horizon before matter-radiation equality. Therefore, we integrate over all the DWs that reenter the horizon between $t_\text{eq}$ and the time at which the DWs form, $H(T_f)=m_a$. 
If the dark matter from the DWs is to account for all the observed DM then the majority of the walls should indeed have entered the horizon well before matter-radiation equality. This is not required if instead the DM from domain walls is only a small fraction of the total DM. It is however straightforward to extend our expressions to this more general case.

Putting all these ingredients together we find the amount of dark matter from the walls to be given by
\begin{eqnarray}
	\rho_\text{dm}^\text{dw} (t_0)= \int_{N_\text{eq}}^{N_\text{osc}}	\beta (N_*)    \, \rho_\text{dw}\, \parfrac{a(H_*)}{a_\text{0}}^3 d N_* 
    \label{eq: Dark Matter from  Domain Walls}
\end{eqnarray}
where the subscript $\text{osc}$ denotes quantities evaluated at the time of DW formation.
In practice, we further change the integration variable from the e-fold number $N_*$ to the dimensionless quantity $x=T_*/T_\text{eq}$ where $T_*=T(H_*)$ is the temperature at the horizon crossing time of the scale $k_*$.

\subsection{Primordial Black Holes}

The domain walls keep entering the horizon with larger and larger masses. In the spherical wall approximation, a DW that enters the Hubble sphere has a mass
\begin{eqnarray}
    M_\text{dw}= 4 \pi \sigma/H^2 + 4\pi \Delta V/(3H^3)
    \label{eq: PBH mass} \,.
\end{eqnarray}
It is clear that its associated Schwarzschild radius, $r_s=2 G M_\text{dw}$, grows faster than the horizon  $r_H=1/H$ and so eventually becomes of the size of the Hubble radius. At that point,  the closed DW entering the Hubble sphere will also be inside its Schwarzschild radius and thus becomes a Primordial Black Hole (PBH) for an observer outside the DW (as in~\cite{Deng:2016vzb}).

PBHs can in principle form even earlier; if during the DW collapse there is no significant energy loss nor large angular velocities the DW will eventually enter its Schwarzschild radius.
For spherical walls the formation of a PBH is known to be very efficient \cite{Deng:2016vzb}. However, in the diffusion mechanism that we discuss here, we do not expect the walls to be spherical and there could be spinning-up processes and energy loss mechanisms such as scalar and gravitational wave emission halting the formation of the PBH.  A detailed study that characterizes this dissipation is necessary.
Therefore,  
to assess whether a PBH is formed it is convenient to define the so-called figure of merit, $\text{FoM} \simeq r_\text{Sch}/r_\text{dw}$ \cite{Vachaspati:2017hjw, Ferrer:2018uiu}, that characterizes how close to its Schwarzschild radius is a Hubble-sized domain wall of radius $r_\text{dw}=1/H$.
The figure of merit grows as
\begin{eqnarray}
    \text{FoM} = \frac{8 \pi \sigma_{\text{dw}} G}{H}  + \frac{8 \pi \Delta V G}{3H^2} \, .
\end{eqnarray}
While the \text{FoM} has been computed for a spherical surface, even an aspherical wall with asphericities as large as the Hubble radius is expected to collapse to a PBH if $\text{FoM}>1$, since the Schwarzschild radius would be itself comparable to the largest extension of the re-entering surface. Therefore, we will consider the time of PBH formation $t_\text{PBH}$ to be the time at which the FoM $\simeq {\cal O}(1)$ (the most conservative case) but also show results for FoM $\simeq {\cal O}(0.1)$. As we show in Appendix \ref{App: PBH dominated region}, in all the cases of interest we find that the majority of the dark matter from the walls is in the form of PBHs.

To compute the PBH abundance we just need to perform the same integral as in Eq. \ref{eq: Dark Matter from  Domain Walls} but restricted to the times $t\geq t_\text{PBH}(\text{FoM})$, where we have explicitly included the dependence on the choice of the figure of merit. In summary, the energy density of PBH today is
\begin{eqnarray}
	\rho_\text{PBH} (t_0)&=& \int_{N_\text{eq}}^{N_\text{max}}	\beta (N_*)    \, \rho_\text{dw}\, \parfrac{a(H_*)}{a_\text{0}}^3 d N_* 
 \label{eq: rho_pbh}
\end{eqnarray}
where $N_\text{max} = \min \left[N(m_a=H),N_{t=t_\text{PBH}}\right]$. 

Besides the overall abundance, another critical quantity is the PBH mass distribution $d f_\text{PBH}(M_\text{PBH})/d \log M_\text{PBH}$ which is currently subject to strong constraints over a wide range of masses. This distribution can be obtained from Eq. \ref{eq: rho_pbh} by noting that in general the PBH abundance is given by
\begin{equation}
    f_\text{PBH}\equiv \frac{ \rho_\text{PBH}}{\rho_\text{DM}} = \int_{M_\text{PBH}^\text{min}}^{M_\text{PBH}^\text{max}} \frac{d f_\text{PBH}(M_\text{PBH})}{d \log M_\text{PBH}} d \log M_\text{PBH} \label{eq: OmegaPBH log decomposition} \, .
\end{equation}
The upper and lower limits of integration are calculated from the relation between $N_*$ and $M_\text{PBH}$, using Eqs. \ref{eq: N_* (T_*)} and \ref{eq: PBH mass}, and define a maximal and a minimal PBH mass. The minimal mass corresponds to the PBHs that are formed at the earliest time ($t_\text{PBH}$). We also cut the mass distribution at a maximal mass, corresponding to walls that enter at matter-radiation equality. We have checked that the contribution to the total abundance from the PBHs that form after that time is negligible in all the cases considered in this work.

\section{Isocurvature Constraints and Misalignment Contribution }
\label{sec: Isocurvatures}

To form DWs that re-enter the horizon before matter-radiation equality, we necessarily rely on sizeable isocurvature perturbations of the axion field during inflation. However, these perturbations are on small (non-cosmological) scales and are thus unconstrained by observations.
At CMB scales, the isocurvature perturbations are smaller, of order $\sigma_{\theta}(N_\text{cmb})\simeq H/(2\pi f)$, because there is less time for diffusion. Nonetheless, Planck data~\cite{Planck:2018jri} still places strong bounds on the model. 

Following \cite{Kobayashi:2013nva}, the ratio of isocurvature to adiabatic perturbations in the CMB is given by
\begin{eqnarray}
\alpha_\text{iso} =	\frac{P_\text{iso}(f_a,H_I, \theta_i^\text{eff})}{P_\zeta}=\frac{1}{P_\zeta}\left[ \frac{\Omega_a^\text{mis}(\theta_i^\text{eff})}{\Omega_\text{dm}}  \frac{\partial \log \Omega_a}{\partial \theta_i^{\text{eff}}}  \sigma_{\theta}(N_\text{cmb}) \right]^2 <0.038 
\label{eq:isocurvature formula}
\end{eqnarray}
where we have imposed the latest CMB constraint on $\alpha_\text{iso}$~\cite{Planck:2018jri}.
The previous expression includes the enhancement effects of the anharmonicities when the initial angle $\theta \gtrsim 1$  and simplifies to the more conventional expression $\alpha (\theta_i\ll 1)= 1/P_\zeta \left( (\Omega_a^\text{mis}/\Omega_\text{dm}) \, H_I/( \theta_i^\text{eff} \pi f_a) \right)^2$  for $\theta_i \ll 1$.  Furthermore, given that we are interested in exploring regions where $\theta_i \gtrsim 1$, we do not rely on the analytical approximations valid for $\theta_i \ll 1$ to calculate the misalignment abundance, and instead directly evaluate the axion abundance by solving its equation of motion and following the steps outlined in appendix S11 of \cite{Borsanyi:2016ksw} (see appendix \ref{app: Numerical evaluation of the Misalignement contribution} for more details).

In Eq. \ref{eq:isocurvature formula} we have used the quantity $\theta_i^\text{eff}$, defined by
\begin{eqnarray}
\label{eq:thetai}
    \theta_i^\text{eff}=   \sqrt{ \theta_i^2 + \sigma_{\theta}(N_\text{osc})^2} \,.
\end{eqnarray}
Indeed, while we are dealing with a pre-inflationary PQ-breaking scenario where there is a homogeneous value of $\theta_i$ after inflation, quantum diffusion can be larger than the initial angle $ \sigma_{\theta}(N) \gtrsim \theta_i $, thereby dominating the root mean square $\sqrt{\left<\theta\right>^2}$ of the field. This is particularly relevant in this work because we rely on a non-negligible diffusion to later form the DWs. We thus evaluate the diffusion at the e-fold $N_\text{osc}$, corresponding to the scale that re-enters the Hubble sphere at the time that the axion mass becomes comparable to the Hubble scale, and thus the matter-like oscillations start.~\footnote{ We take this time to be at $T \simeq 5$~GeV for the QCD axion and $T \simeq \sqrt{m_a M_p}$  for the ALP.  A more precise determination of $T$ as a function of $f_a$ can in principle be implemented. However, given that it only affects the e-fold number logarithmically we take these values as representative.} Perturbations that re-enter before this time (thus at $N>N_\text{osc}$) are diluted more rapidly, as radiation, and thus do not contribute significantly to the determination of $\theta_i^\text{eff}$. Note that our expression for the isocurvature bound is more conservative than previous estimates~\cite{Lyth:1991ub,Hertzberg:2008wr,Marsh:2015xka}.

\section{Results}
\label{sec: Results}

So far we derived the general expressions for the dark matter, PBHs and isocurvatures that come from the DWs. We will now explore two specific and well-motivated examples: the QCD axion and a general Axion-Like Particle (ALP) scenario. 

\subsection{QCD axion}

In the case of  the QCD axion, the axion mass is temperature dependent and given by $m_a(T) = \sqrt{\chi(T)}/f_a$
with $\chi(T)$  the QCD topological suspectability that we take from \cite{borsanyi2016latticeqcdcosmology}.

We anticipate that, whenever the dark matter coming from the walls is sizable, the majority of it is in the form of PBHs (see Appendix \ref{App: PBH dominated region}). We found this to be a common result even when moving beyond the QCD axion scenario. Therefore,
we start by discussing the PBH mass distribution and the associated bounds, then isocurvature perturbations, and finally the dark matter abundance.

\begin{figure}[t]
\centering
        \includegraphics[width=0.75\textwidth]{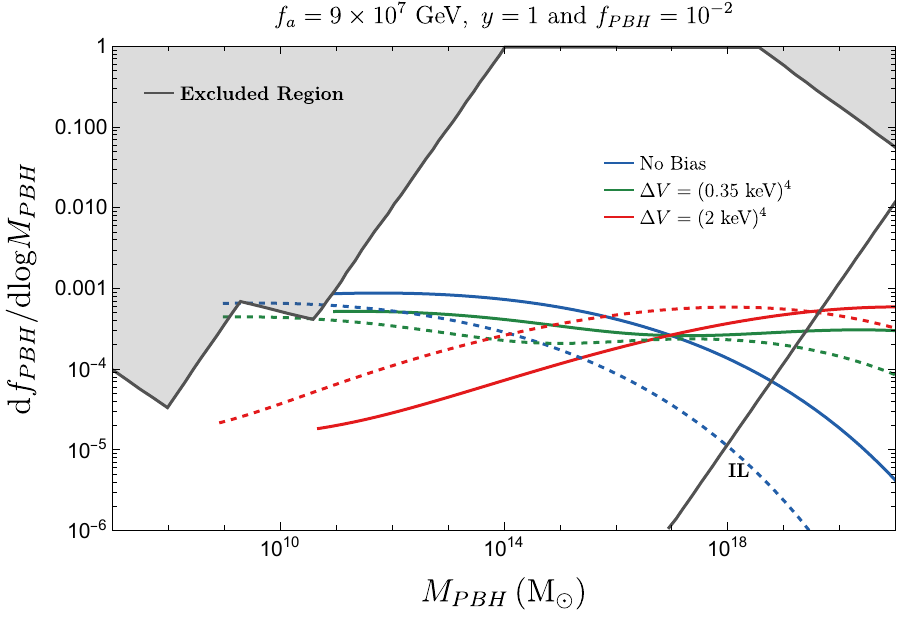}
        \caption{QCD axion case. PBH mass distribution plot for fixed $f_{\text{PBH}} = 10^{-2}$, $f_a= 9 \times 10^7\,$GeV, $y= 1$ and different values of the bias $\Delta V$.  The straight (dashed) lines represent the PBH mass distribution for FoM$=1$ (0.1). The parameters chosen satisfy isocurvature constraints.  In the grey region, we show the constraints on PBH mass fraction taken from \cite{Carr_2022}.
        }
        \label{fig:dLogfdLogM QCD PBH}
\end{figure}

 \paragraph{PBH mass distribution:}
In Fig. \ref{fig:dLogfdLogM QCD PBH} we plot the PBH mass distribution, defined in  Eq. \ref{eq: OmegaPBH log decomposition}, as a function of the PBH mass, in solar masses, for $f_a=9 \times 10^7\text{ GeV}$ and $y = 1$. The value of $\theta_i$ has been fixed by imposing $f_\text{PBH}=10^{-2}$ and the figure of merit for PBH formation has been chosen to $\text{FoM}=1$ (straight lines) and $\text{FoM}=0.1$ (dashed lines). 
In blue we show the results without bias, whereas the green and red curves correspond to the case with bias parameter $\Delta V^{1/4}$ of $0.35$ and $2\text{ keV}$ respectively.\footnote{The axion solution to the strong CP problem is preserved as long as $\Delta V^{1/4}\lesssim 100~\text{keV}$.} 
All curves satisfy the isocurvature constraints that we will discuss afterward. For this particular value of parameters, the amount of non-PBH dark matter from the DWs is comparable to that in the form of PBHs for the blue and green curves.

Let us highlight two prominent features of the PBH mass distribution shown in Fig.~\ref{fig:dLogfdLogM QCD PBH}. First, PBHs start forming very late, below MeV temperatures, and so have very large masses, above $10^9 \, M_\odot$.  
Smaller masses, require larger $f_a$ and are in tension isocurvature and PBH constraints (in grey we show bounds from CMB dipole, large-scale structure, dynamical friction and galaxy tidal distortions, all taken from \cite{Carr_2022}).~\footnote{We have not included PBH constraints coming from $\mu$-distortions as this mechanism is non-Gaussian and does not rely on adiabatic perturbations, conditions under which those bounds are applicable.
 It is important to mention that the constraints typically assume a monochromatic mass function, whereas the mechanism considered here predicts a broad mass function. However, we still expect these constraints to apply to our scenario at the order of magnitude level. In this regard, we have chosen in Fig. \ref{fig:dLogfdLogM QCD PBH} the maximal value of $f_a$ that is compatible with PBH constraints in the bias-less case.}
Second, the mass distribution is rather flat over several epochs of PBH mass. This is related to the logarithmic dependence of $N_*$ on the temperature (see Eq. \ref{eq: N_* (T_*)}) which translates into a relatively slow suppression of the DW contribution on large scales. 

The addition of a bias between minima has two main effects. First, the DWs become more massive and so PBHs start forming earlier on. Second, the peak of the distribution moves to larger masses because the vacuum energy inside the DWs counterbalances the suppression in the number of DWs entering the horizon for some time. For large enough bias, we find that the peak position depends mostly on $\theta_i$ and $y$ and becomes almost independent of the bias parameter.
Eventually, the exponential tail of the distribution dominates over this effect at large masses and cuts off the spectrum.
For $\text{FoM}=1$ the peak of the distribution is however inside the so-called \textit{incredulity region} where there is on average less than one object within the observable universe and therefore where our continuous PBH mass distribution stops being an appropriate description. 

Note also that the (dashed and solid) curves in Fig.~\ref{fig:dLogfdLogM QCD PBH} are cut off at small PBH masses. This is because we only consider PBH to form when the condition $t \gtrsim t_\text{PBH}$ (see Section \ref{sec: Dark matter and Primordial Black Holes}). Relaxing the PBH formation criterion to $\text{FoM}\geq 0.1$ (dashed lines) moves the curves to lower masses (smaller $t_\text{PBH}$).

\begin{figure}[t]
\centering
    \includegraphics[width=0.75\textwidth]{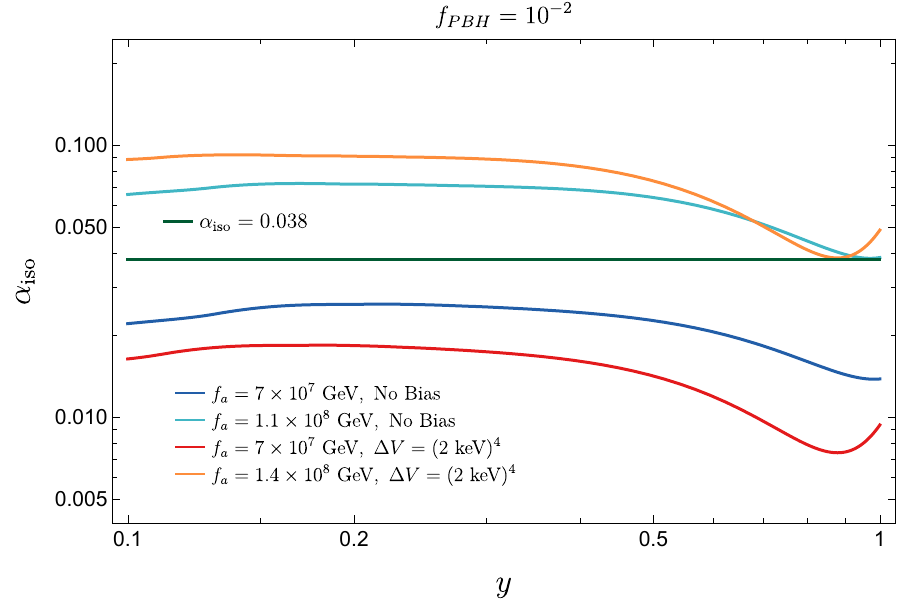}
        \caption{Ratio of isocurvature to adiabatic perturbations at CMB scales for the QCD axion as a function of the parameter $y$. We have fixed $\theta_i$ such that $f_{\text{PBH}} = 10^{-2}$. The green line denotes the CMB constraint on $\alpha_\text{iso}$ \cite{Planck:2018jri}. The dark and light blue curves are biasless and have  $f_a=7 \times 10^7$  and $f_a=1.1 \times 10^8$, respectively. The red and orange curves have a bias parameters curve of $(2 \ \text{KeV})^4$. We have fixed the figure of merit to one in all curves.}
        \label{fig:QCD isocurvatures plot}
\end{figure}

\paragraph{Isocurvature Perturbations:}
In Fig. \ref{fig:QCD isocurvatures plot} we plot the ratio of isocurvature to adiabatic perturbations in Eq. \ref{eq:isocurvature formula}, as a function of $y=H_I/f$, the parameter that controls the amount of diffusion, for different values of the QCD axion decay constant $f_a$ and the bias parameter $\Delta V$. The initial misalignment angle $\theta_i$ has been chosen such that the abundance of PBH is $f_\text{PBH} = 10^{-2}$.
Let us first discuss the cases without bias.
For  $f_a \simeq 7 \times 10^{7} \, \text{GeV}$ (dark blue) the isocurvature spectrum is below the current upper bound of $0.038$ \cite{Planck:2018vyg} for any value of $y$. As we increase $f_a$ the isocurvature spectrum increases. The maximal value of $f_a$ compatible with observations  is $f_a \simeq 1.1 \times 10^8 \, \text{GeV}$ for $y \simeq  1$ (light blue).

In the presence of a bias term, the allowed window for the decay constant is widened. This is because, due to the bias, the DWs have more energy and so to get the same amount of DM as in the biasless case one needs fewer DWs and also fewer isocurvatures. For example, for $\Delta V= (2~\text{keV})^4$ the maximal value of $f_a$ allowed by observations increases up to $f_a = 1.4 \times 10^8 \, \text{GeV}$ for $y \simeq 0.9$.  A larger bias parameter further alleviates the isocurvature bound. However, it moves the peak of the PBH distribution beyond the incredulity limit and so outside the regime of validity of the analysis, as we have discussed before.

\begin{figure}
\centering\includegraphics[width=0.49\textwidth]{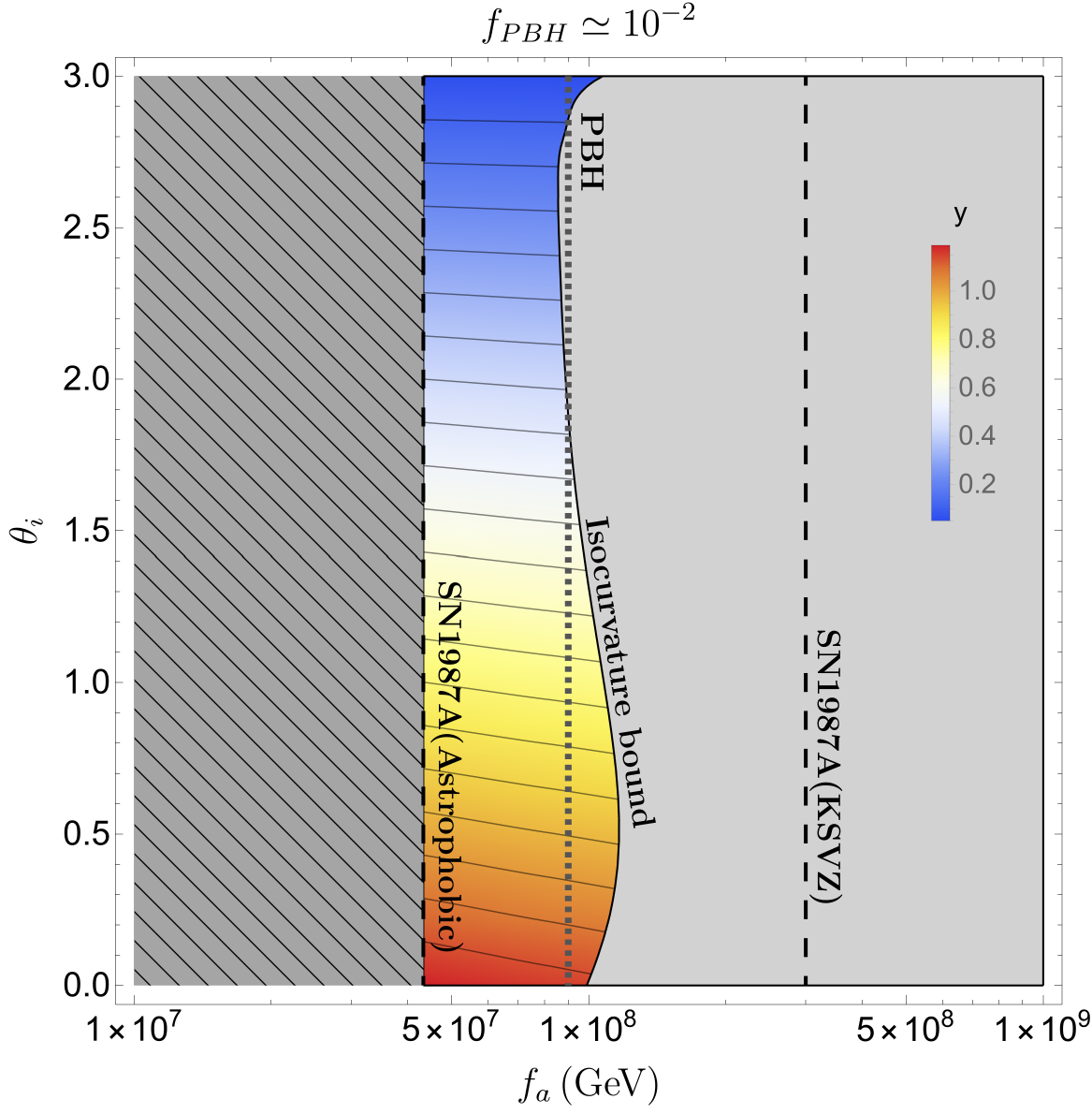}
\,
\includegraphics[width=0.49\textwidth]{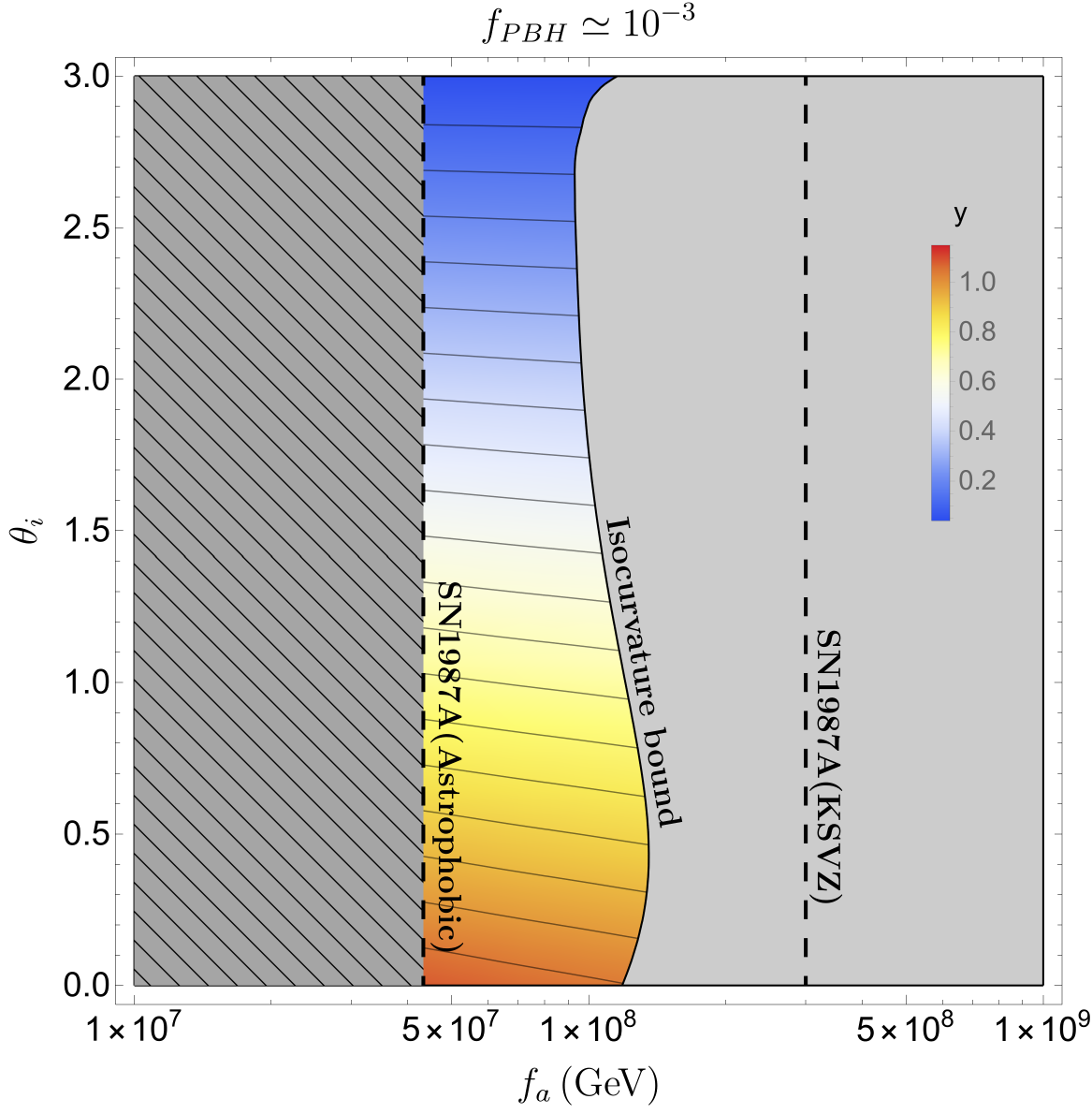} 
	\caption{QCD axion parameter space with $\Delta V=0$, for two values of the fraction of DM in PBHs: $f_\text{PBH}=10^{-2}$ (left panel), $f_\text{PBH}=10^{-3}$ (right panel). The colored region is allowed by isocurvature constraints~\cite{Planck:2018jri}. PBH bounds (dashed line,  $\text{FoM}=1$) set an upper limit on $f_a$ (dotted vertical line) in the case $f_\text{PBH}=10^{-2}$, whereas the bound becomes significantly relaxed (and larger than $10^9$) for $f_\text{PBH}=10^{-3}$. The SN1987A bound imposes a lower limit on $f_a$: given its astrophysical uncertainties (see e.g. \cite{Bar:2019ifz}), as well as its dependence on the axion coupling to nucleons, we show two representative lower limits on $f_a$ corresponding to recent estimates~\cite{Carenza:2019pxu} for the KSVZ model and the astrophobic axion \cite{DiLuzio:2017ogq,DiLuzio:2024vzg}.}
	\label{fig:QCD Axion Parameter Space}
\end{figure}

\paragraph{Dark Matter contribution:}
Finally, in Fig. \ref{fig:QCD Axion Parameter Space} we combine all the information above and show the range of parameters $y, \theta_i, f$ for which the dark matter from the domain walls (see Eq. \ref{eq: Dark Matter from  Domain Walls}) accounts for one percent (left panel, or $0.001$ in the right panel) of the observed dark matter abundance (gradient colored region) while being compatible with isocurvature constraints.

Importantly, in these regions, dark matter from DWs is mostly in the form PBHs. Therefore, large contributions to the total DM abundance are strongly constrained by the PBH bounds. 
In this particular case, such bounds imply a constraint $f_a \lesssim 9 \times 10^{7}$ GeV for $f_\text{PBH} \simeq 10^{-2}$ (see Fig.~\ref{fig:dLogfdLogM QCD PBH}). We note however that this bound should be taken as an order of magnitude estimate, because of the associated uncertainties in the PBH bounds. Requiring a larger fraction of PBH DM would further strengthen the bound whereas a lower fraction, even if by just a factor of two, can significantly alleviate the upper limit to values above $10^9$ GeV. In the the right panel of Fig.~\ref{eq: Dark Matter from  Domain Walls} the case with $f_\text{PBH}=0.001$ is shown, which is not subject to PBH DM bounds for $f_a\lesssim 3 \times 10^{10}$ GeV.

A $y$-dependent upper bound on $f_a$ from the overproduction of isocurvature perturbations in Eq. \ref{eq:isocurvature formula}, as well as the usual lower bound on $f_a$ from observations of SN1987A are also shown.
This latter bound, besides the possibly significant astrophysical uncertainty (see also discussion in~\cite{Bar:2019ifz}), exhibits a mild model dependence, due to the model-dependent couplings of the QCD axion to nucleons. We present two representative cases: the (strongest) bound assuming a standard KSVZ-like coupling to nucleons~\cite{Carenza:2019pxu} and a weaker bound corresponding to the so-called astrophobic axions \cite{DiLuzio:2017ogq,DiLuzio:2024vzg} where the coupling to nucleons is maximally suppressed. 
On the other hand, isocurvature perturbations place an upper limit on $f_a$ that is weaker than the PBH constraints for most values of parameters in the case of $f_\text{PBH}=10^{-2}$, although it rapidly becomes the dominant upper bound once $f_\text{PBH}$ is smaller. Note also that the isocurvature bound is only very mildly relaxed as $f_\text{PBH}$ decreases.

In Section \ref{sec: Evading Isocurvatures} we will discuss the possibility of evading isocurvature bounds, thereby opening up the viable region of parameter space.

\subsection{Axion-Like Particles}

Let us now consider the case of ALPs. The main differences compared to the previous subsection are: 1) we treat $m_a$ and $f_a$ as independent parameters; 2) we do not include the possible temperature dependence of the ALP mass; 3) we do not impose the SN1987A bound on $f_a$ as we consider ALPs that do not couple significantly to Standard Model particles. 
As we will show, this allows to expand the viable region of parameters and to achieve a larger PBH abundance. 
To simplify the analysis we fix $\theta_i = 1$, although, as we have seen in the previous section, the results are not very sensitive to the choice of $\theta_i$ parameter. 

\begin{figure}
    \centering
    \includegraphics[width=0.75\linewidth]{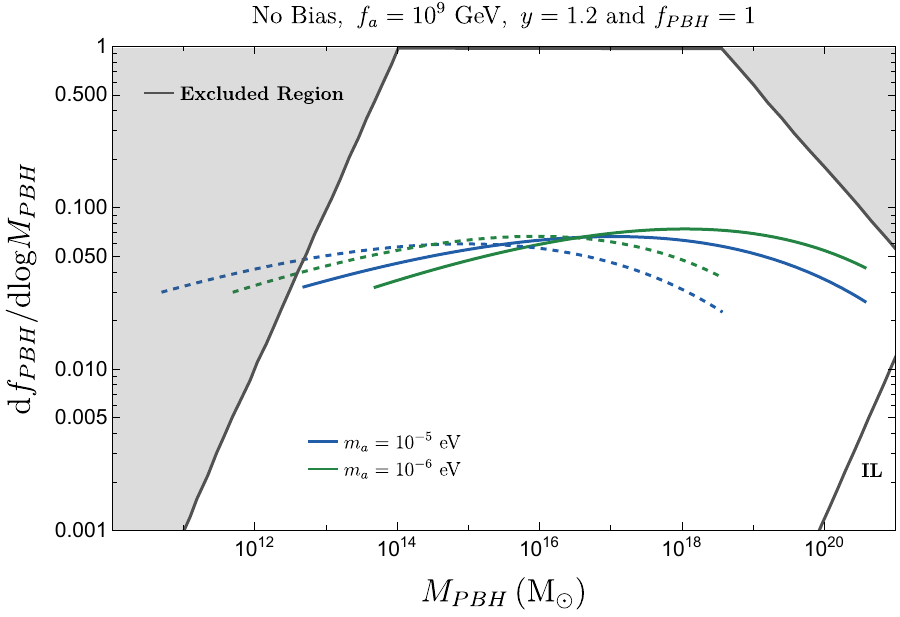}
    \caption{ALP scenario. PBH mass distribution plot for fixed $f_{\text{PBH}} = 1$ and $f_a=10^9$ GeV and two values of the ALP mass.  The straight (dashed) lines correspond to a figure of merit of $1$ ($0.1$). The parameter $y$ has been chosen such that isocurvature constraints are satisfied and $\theta_i$ has been fixed to unity.}
    \label{fig:PBH ALPs large density}
\end{figure}

As in the QCD axion case, we find that the majority of the dark matter from the walls comes in the form of PBH (see Section \ref{App: PBH dominated region}) in all the regions of interest.
In the case of ALPs, there are several regions of phenomenological interest in terms of PBH formation. Here we focus on the region of parameters where the PBH abundance can be large, $f_\text{PBH}={\cal O}(0.1-1)$. Currently, there are only two mass windows where the distribution $d f_\text{PBH}/d\log M_\text{PBH}$ can reach values close to one: the asteroid range, for masses between $10^{-16}M_\odot-10^{-10}M_\odot$ and the stupendously large window, for masses between $10^{14}M_\odot-10^{19}M_\odot$, which surprisingly seems to remain unconstrained \cite{Carr_2022}. However, we find that $f_\text{PBH}\simeq 1$ can be achieved only for stupendously large PBH masses, if isocurvature constraints are to be satisfied. We comment on other interesting phenomenological regions in Section \ref{sec: Evading Isocurvatures} and Fig. \ref{fig:dLogfPBH for high fa} when considering scenarios that evade isocurvature bounds.

\begin{figure}
    \centering
\includegraphics[width=0.7\linewidth]{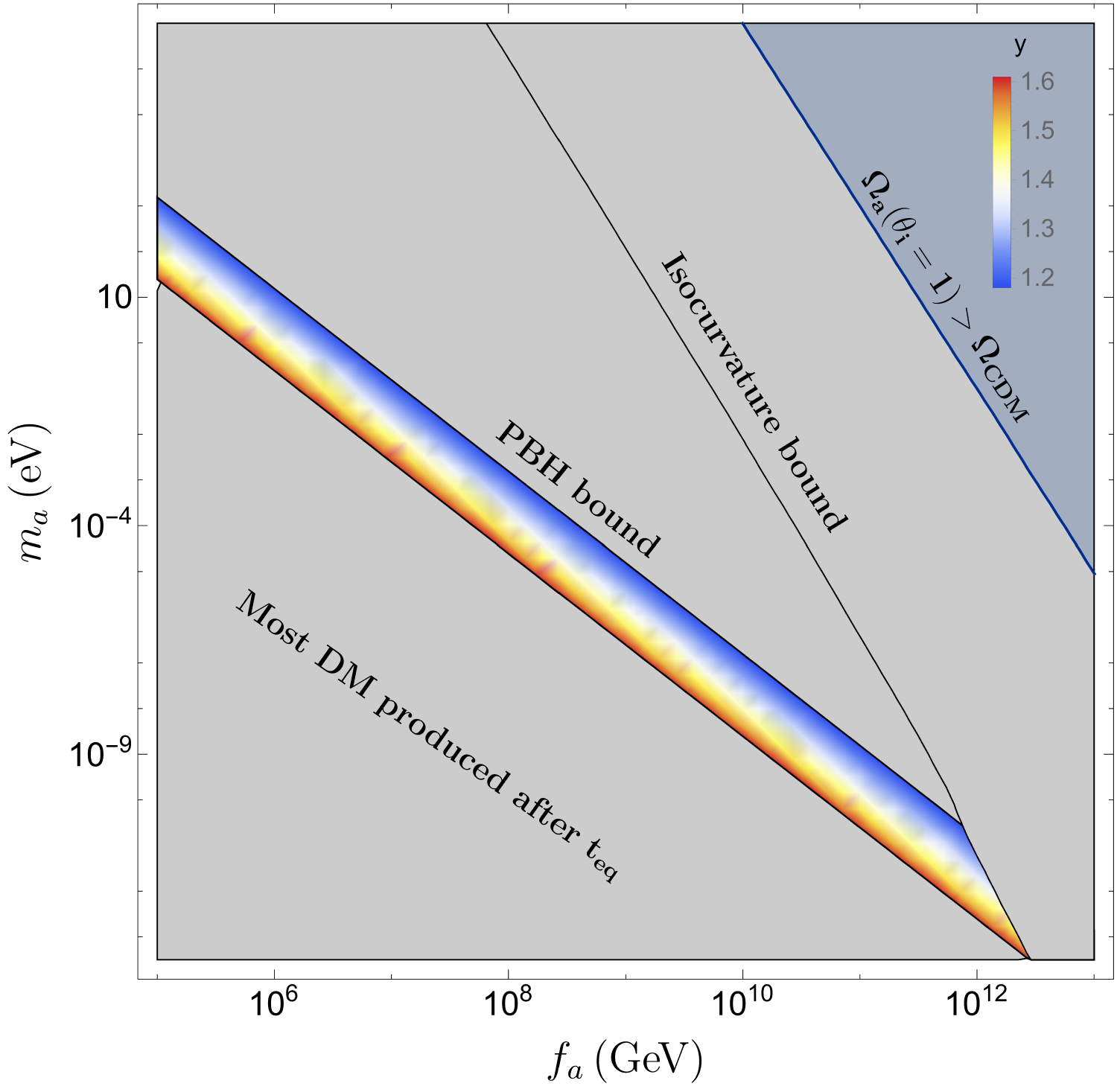}
    \caption{ALP parameter space for a fixed PBH fraction $f_\text{PBH}\simeq 1$ ($\theta_i$ and FoM have been fixed to one). The viable (colored) region is limited (in grey) from above by the isocurvature and PBH constraints and from below by imposing that most DM is formed before equality (also in grey). The blue-shaded region corresponds to the overproduction of DM from the misalignment mechanism. }
    \label{fig:ALPs all DM and xPBH vs xPeak}
\end{figure} 

In Fig. \ref{fig:PBH ALPs large density} we plot the PBH mass distribution for $f_a=10^9$ GeV and $m_a=3\times 10^{-14}$ eV. The parameters are chosen such that the abundance of PBHs can explain all/most of the DM while satisfying isocurvature bounds.  
As for the QCD axion, the distribution is rather broad and ends at small masses when the condition for PBH formation stops being satisfied. The distribution is also cut at large masses, around $10^{20} M_\odot$, corresponding to the mass of the DW that enters at matter-radiation equality.~\footnote{One could in principle extend the distribution to larger masses by extending the integrals in Eq. \ref{eq: rho_pbh} beyond the matter-radiation equality. However, this part of the distribution would lie in the \textit{incredulity region}, where the computation we are performing is not valid. In any case, the contribution from BHs that form after matter-radiation equality to the total abundance is subleading.} We verified that for FoM $=1$, we can have  $f_\text{PBH}\simeq 1$ while satisfying current PBH and isocurvature constraints whereas for FoM $=0.1$ one would need to lower $f_a$ or $m_a$ in order not to be in tension with PBH bounds.

Note however that the values of $f_a, m_a$ for which the PBH can be all the dark matter are not unique. Basically, all the DWs that have a similar tension yield a similar PBH mass distribution. This can be understood from Eq. \ref{eq: rho_pbh} since the dependence of $f_\text{PBH}$ on $m_a,f_a$ comes mostly through the DW tension $\sigma_{dw}\propto m_a f_a^2$. Furthermore, DWs with smaller tension  move the PBH distribution to larger masses and so can also yield $f_\text{PBH} \simeq 1$. However, the DW tension cannot be so small that the peak of the distribution moves to post matter-radiation equality scales.

In Fig. \ref{fig:ALPs all DM and xPBH vs xPeak}, we show the region of parameters $m_a, f_a, y$ (for FoM $=1$) where PBHs can account for all of the dark matter in the window of stupendously large masses.
 The available parameter space is constrained to  $f_a \lesssim 3\times 10^{12}$ GeV, $m_a >10^{-13}$ eV from the isocurvature bound and the fact that most PBHs need to be formed before $t_\text{eq}$. 
 In the same plot, we also show the parameter space where the total dark matter abundance is produced by the standard misalignment mechanism (for $\theta_i=1$). This window corresponds to values of $f_a$ and $m_a$ that are much larger than those achievable through DWs, and that may thus be more challenging to test.

\section{Evading Isocurvatures and PBH bounds}
\label{sec: Evading Isocurvatures}

In the previous section, we found that the isocurvature bound severely limits the viable range of $f_a$, where a significant fraction of DM in PBHs can form, in the QCD axion case. Significant constraints also arise from PBH overproduction, also in the ALP case. In this section, we discuss two possible ways to avoid such bounds.

One possibility is that inflation unfolds in multiple stages at different energies. For example, one can have two stages, one that happens at high energy scales, while the $U(1)$ symmetry is still unbroken, and lasts for a few ${\cal O}(1-10)$ e-folds, and a second stage at lower energy, when the $U(1)$ symmetry is already broken so that the diffusion takes place. In this way, isocurvature perturbations would only be generated at the (smaller) scales that exited the horizon during the second stage and are thus unconstrained by the CMB observations. This idea has recently been explored in the context of cosmic strings \cite{Redi:2022llj}. Similarly, the absence of isocurvature perturbation at the largest scales would also suppress the formation of PBH at later times (larger masses). 

A second example where large-scale isocurvature perturbation can be suppressed is when the axion decay constant $f_a$ varies during inflation (see e.g. \cite{Lyth:1991ub,Gorghetto:2023vqu}). This can be achieved if, for example, the PQ radial mode is light during the epoch at which the CMB scales exited the horizon, and such that its dynamic drives $f_a$ to larger values, thereby suppressing both isocurvature perturbations and PBH formation at CMB scales. At later times, once $f_a$ settles to its low-energy value, diffusion can then become large as described above.

\begin{figure}
    \centering
\includegraphics[width=0.9\linewidth]{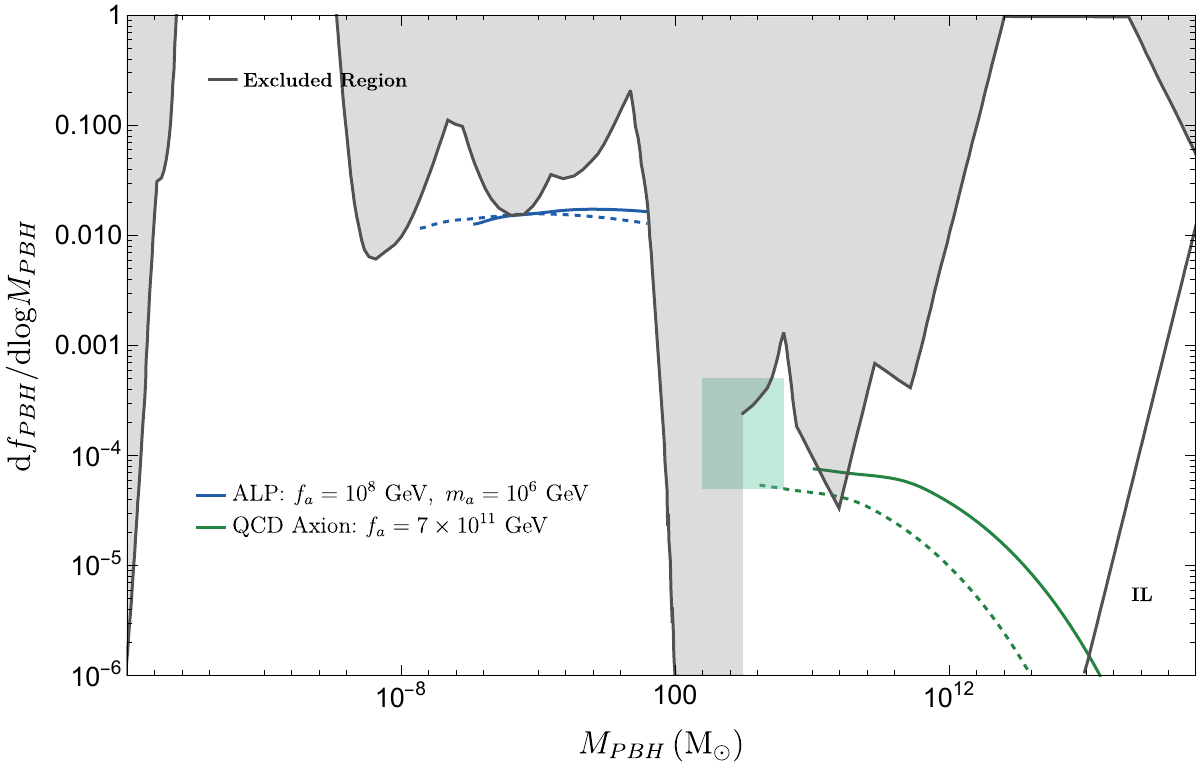}
    \caption{
    Green:  PBH mass distribution for the QCD axion for a fixed $f_{\text{PBH}} = 8\times 10^{-4}$. The green square represents the region where PBHs could help in explaining the early formation of ultracompact objects \cite{Silk:2024rsf}. Blue: ALP PBH mass distribution plot for a fixed $f_{\text{PBH}} \simeq 0.5$. Straight lines represent the PBH mass distribution for a FoM of $1$, while the dashed lines represent the PBH mass distribution for a FoM of $0.1$. The grey area denotes the constraints on $f_{\text{PBH}}$ \cite{Carr_2022}. 
    \label{fig:dLogfPBH for high fa}}
\end{figure}

One can now ask the question, what other regions of phenomenological interest open up once this isocurvature and PBH suppression occurs? In Fig. \ref{fig:dLogfPBH for high fa} we show two examples, one for the QCD axion (in green) and another for an ALP (in blue).
Let us start with the green curves. They correspond to a value of the decay constant such that the standard misalignment mechanism of the QCD axion produces all of the dark matter, i.e. $f_a \sim 7 \times 10^{11}$ GeV. Yet, there is still a subdominant ($f_\text{PBH}\sim 8\times 10^{-4}$), albeit observationally interesting, fraction of DM in the form of PBHs with masses $10^5-10^6 M_\odot$ when FoM$\,=0.1$ (dashed curve).
These abundances and masses (depicted as a green-shaded region) align with those recently discussed in~\cite{Silk:2024rsf}, where PBHs can offer a potential explanation for the early formation of (ultra-compact) galaxies~\cite{JWST}.

Regarding the blue curves, it corresponds to the scenario of an ALP with $f_a=10^8$ GeV and $m_a=10^6$ GeV that yields PBHs with masses between roughly $10^{-7}-10^{-6}M_\odot$ and  $100 M_\odot$ and that amount to roughly half of the total dark matter abundance. In both the case with FoM$\,=1$ (straight line) and with FoM$\,=0.1$ (dashed line), we have chosen the maximal PBH mass to be lower than the maximal allowed by the CMB accretion bound \cite{Ali-Haimoud:2016mbv}. In terms of the two mechanisms mentioned at the beginning of this section, the maximal PBH mass amounts to choose the scale at which isocurvatures, and so PBHs, disappear.

\section{Conclusions}
\label{sec: Conclusions}

In this work, we took a fresh look into the mechanism of domain wall formation from inflationary perturbations, initially proposed and discussed in \cite{Linde:1990yj,Lyth:1991ub}. We discussed some novel interesting phenomenological aspects, namely, that the mechanism can give rise to dark matter and that it comes with a sizeable fraction of PBHs. Interestingly, this is the case even in axion models where the domain wall number is unity.

The DWs have their origin in the inflationary diffusion of a scalar field with a discrete symmetry, such as the QCD axion or Axion-Like particles. In the regions of space where the diffusion is able to move the field over the maximum of the low-energy scalar field potential, the field will oscillate around a different minimum after inflation and be surrounded by a closed domain wall. For the number of domain walls formed to be sizeable, this typically requires the Hubble scale during inflation to be comparable to the axion decay constant $H_I \sim f_a$.
The rate at which DWs enter the horizon after inflation is related to the probability distribution dictated by the stochastic diffusion (see Eq. \ref{eq: probability of forming walls}); at later times fewer and fewer DWs enter the horizon reflecting the fact that the largest scales in our universe underwent less diffusion.

Upon horizon crossing, the DWs collapse and either decay into wall constituents or, if massive enough, form a PBH. In the regions of phenomenological interest, the Schwarzschild radius of the DW is already larger than the Hubble horizon at the time most of the DWs enter the horizon and so most DWs collapse into PBHs. We found the resulting PBH mass distribution to be, in most cases, rather flat and to cover a wide range of PBH masses. This relative flatness is an interesting feature of the mechanism that differs significantly from the PBHs that are formed from critical collapse and from the annihilation of DW networks formed after inflation. 

We then studied the implications of this mechanism for QCD axion models and for general ALPs. 
For the QCD axion, both the observational bound on isocurvature perturbations and the bounds on the PBH mass distribution place a strong upper limit on the axion decay constant $f_a$. 
To have a viable parameter space that satisfies both the latter upper limits on $f_a$ and the lower limit from SN1987A, the PBH fraction needs to be below $f_\text{PBH} \simeq {\cal O} (10^{-2})$. As the PBH fraction decreases, a viable region of parameter space opens for $f_a \simeq 10^8$ GeV. The resulting PBHs are then forced to have \textit{stupendously large} masses, above $10^{11}M_\odot$ (see Fig. \ref{fig:dLogfdLogM QCD PBH}). As an example, for $1$\% of the DM to be in the form of PBHs then $f_a \simeq 10^8$~GeV which is still within the uncertainties of the astrophysical modeling of SN1987A, and within the region allowed in models where the QCD axion coupling to nucleons is suppressed (see summary plot in Fig. \ref{fig:QCD Axion Parameter Space}). Furthermore, this range of $f_a$ is the target of QCD axion direct detection experiments such as IAXO \cite{IAXO:2025ltd}. Additionally, a detectable thermal population of QCD axions is also expected for these values of $f_a$~\cite{Ferreira:2018vjj,Ferreira:2020bpb,Notari:2022ffe}, in models where the coupling to SM particles is not suppressed.

Note, however, that as discussed in Section \ref{sec: Evading Isocurvatures}, some of the constraints can be significantly alleviated if isocurvatures are erased at the largest scales, as explored in recent literature \cite{Redi:2022llj,Gorghetto:2023vqu}. This suppression opens up a few other possibilities, for example: having dark matter from the standard axion misalignment with $f_a= 7 \times 10^{11}$ GeV, and a subdominant abundance of PBHs at masses around $10^{5}-10^{6}M_\odot$ (see Fig. \ref{fig:dLogfPBH for high fa}), that might be enough to explain the puzzling observations of compact objects at high redshifts \cite{Silk:2024rsf}.

Finally, we performed a similar analysis for ALPs. The PBH mass distribution shares several features with the QCD axion case. However, the case of ALPs is less constrained. In particular, we found that it is possible to have all the dark matter (or at least an order one fraction) in the form of PBH with masses above $10^{13}$ solar masses (assuming that such large fractions are observationally viable). In terms of fundamental parameters,  the ALP mass and decay constant that can yield such large fractions, shown in Fig. \ref{fig:ALPs all DM and xPBH vs xPeak}, and are much lower than the typical prediction of the misalignment mechanism.

For the future, it would be rather interesting to think of further observational probes of PBHs with stupendously large masses. One possibility might be through $\mu$-distortions \cite{Deng:2021edw}.  This is a somehow exotic and surprising window which, however, to our knowledge is not yet well constrained observationally.

\section*{Acknowledgements}
We would like to thank Oriol Pujolàs for discussions in the preliminary stages of the project.
RZF and MF acknowledge the financial support provided through national funds by FCT - Fundação para a Ciência e Tecnologia, I.P., with DOI identifiers   10.54499/2023.11681.PEX, \\   10.54499/2024.00252.CERN, 
10.54499/UIDB/04564/2020,
10.54499/UIDP/04564/2020 and 10.54499/2022.03283.CEECIND/CP1714/CT0003 and by the project 10.54499/2024.00249.CERN funded by measure RE-C06-i06.m02 – ``Reinforcement 
of funding for International Partnerships in Science, Technology and Innovation'' of 
the Recovery and Resilience Plan - RRP, within the framework of the financing contract signed between the Recover Portugal Mission Structure (EMRP) and the Foundation for Science and Technology I.P.
(FCT), as an intermediate beneficiary.
The work of F.R.~is supported by the grant RYC2021-031105-I from the Ministerio de Ciencia e Innovación (Spain).

\appendix

\section{Numerical evaluation of the Misalignment contribution}
\label{app: Numerical evaluation of the Misalignement contribution}

 \begin{figure}
 \centering
\includegraphics[width=.495\textwidth]{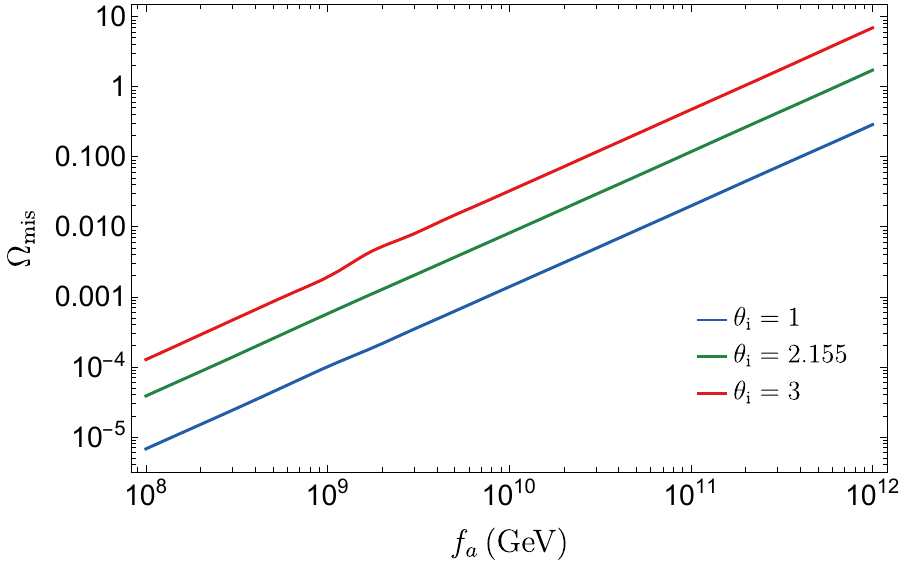} \, \includegraphics[width=.48\textwidth]{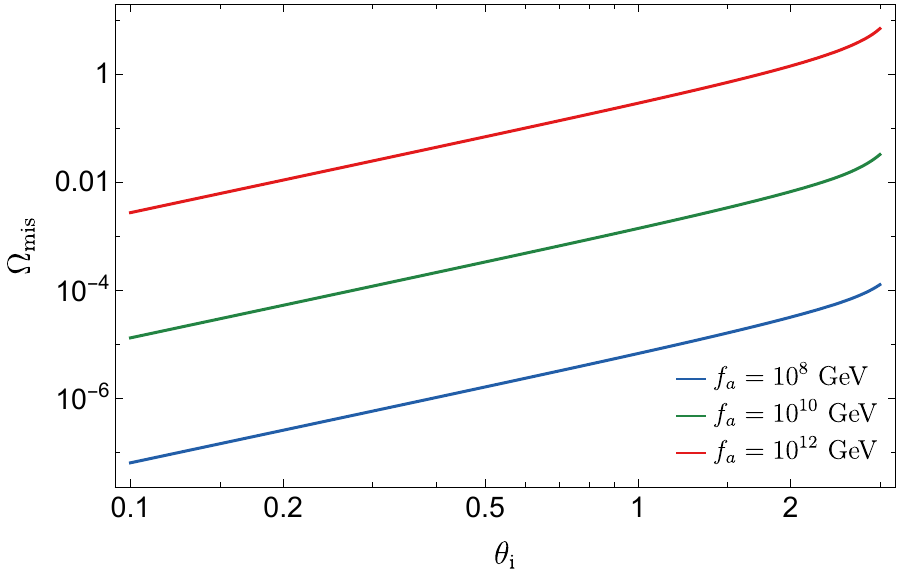} 
	\caption{
    (Left) $\Omega_\text{mis}$ as function of $f_a$ for different values of the initial misalignment angle, $\theta_i$. (Right) $\Omega_{mis}$ as function of $\theta_i$ for different values of the decay constant, $f_a$.}
	\label{fig:Omega_Mis as function of fa and theta}
\end{figure}

In this appendix, we detailed the calculation of the misalignment contribution to dark matter that enters in the computation of the isocurvatures in Eq. \ref{eq:isocurvature formula}.  We are interested in initial misalignment angles $\theta_i$ that are $\gtrsim 1$ and so cannot rely on the analytical approximations valid for small initial angles. Therefore, we computed the misalignment contribution numerically following \cite{Borsanyi:2016ksw}.  

In practice, we solve the Klein-Gordon equation
for the scalar field, 
\begin{eqnarray}
    \frac{d\theta^2}{dT^2} + \left[ 3H(T) \frac{dt}{dT} - \frac{d^2t}{dT^2}/\frac{dt}{dT} \right] \frac{d\theta}{dT} + m_a^2(T) \left( \frac{dt}{dT} \right)^2 \sin{\theta} = 0 \, ,
\end{eqnarray}
in the temperature $T$, instead of time $t$, as a variable.
We start the simulation at temperatures $T_i \gtrsim T_{osc}$, before the onset of matter-like defined by $m_a(T_{osc}) = 3H(T_{osc})$. We take as initial conditions, $\theta(T_i) = \theta_i$ and vanishing first derivative, $\dot{\theta}(T_i)=0$. In the relation between time and temperature, we have included the precise value of the effective number of (entropic) relativistic degrees of freedom, $g_s(T)$ and $g_\rho(T)$, taken from \cite{Borsanyi:2016ksw}.  

We then run the simulation until $ T=0.4 T_\text{osc}$. From the conservation of the number of particles within a comoving volume, we can then calculate the relic abundance of axions today to be,
\begin{eqnarray}
    \frac{n_{a,0}}{s_0}=\frac{n_a(T)}{s(T)}= \frac{45}{2\pi^2}\frac{f_a^2}{m_a(T)g_s(T)T^3} \left[ \frac{1}{2} \left( 
\frac{d\theta}{dT}/\frac{dt}{dT} \right)^2 + m_a(T)(1-\cos{\theta}) \right] \, ,
\end{eqnarray}
where $n_a(T)$ and $s(T)$ are the axion number density and total entropy density, respectively, and the subscript $0$ denotes their values today given, respectively by
\begin{eqnarray}
    s_0 &=& \frac{2\pi^2}{45} \left(2T_\gamma^3 + 6\frac{7}{8}T_\nu^3\right)=\frac{2\pi^2}{45}\frac{43}{11}T_\gamma^3 \, ,\\
    n_{a,0}&=& \frac{\rho_{a,0}}{m_a} = \frac{\Omega^\text{mis}_{a} \rho_\text{crit}}{m_a} 
\end{eqnarray}
where $T_\gamma=2.725$ K is the CMB temperature and  $\rho_\text{crit} \equiv 3H_0^2M_{pl}^2$ is the critical energy density. In computing $n_a(T)/s(T)$ we have averaged over oscillations in the range $0.9-0.4T_{osc}$. 
In Fig. \ref{fig:Omega_Mis as function of fa and theta} we plot the resulting misalignment abundances $\Omega_\text{mis}$ as a function of the decay constant, $f_a$, and the initial misalignment angle, $\theta_i$, respectively.

\section{PBH dominated region \label{App: PBH dominated region}}

\begin{figure}
    \centering
\includegraphics[width=0.455\textwidth]{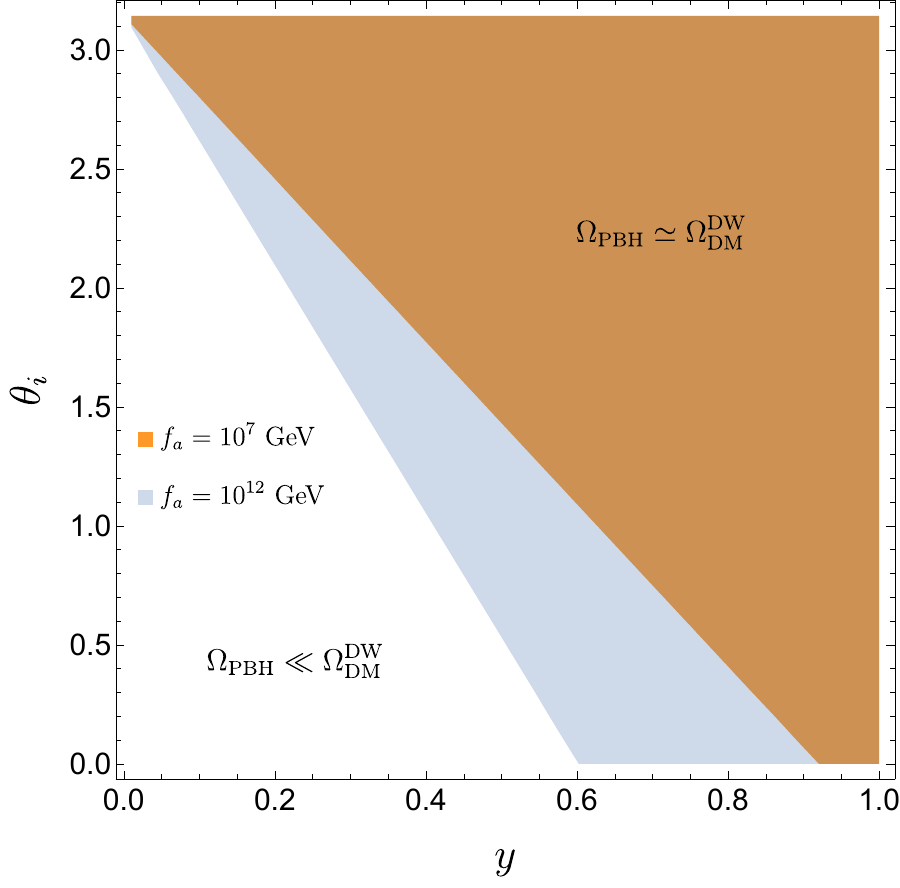}
\includegraphics[width=0.464\linewidth]{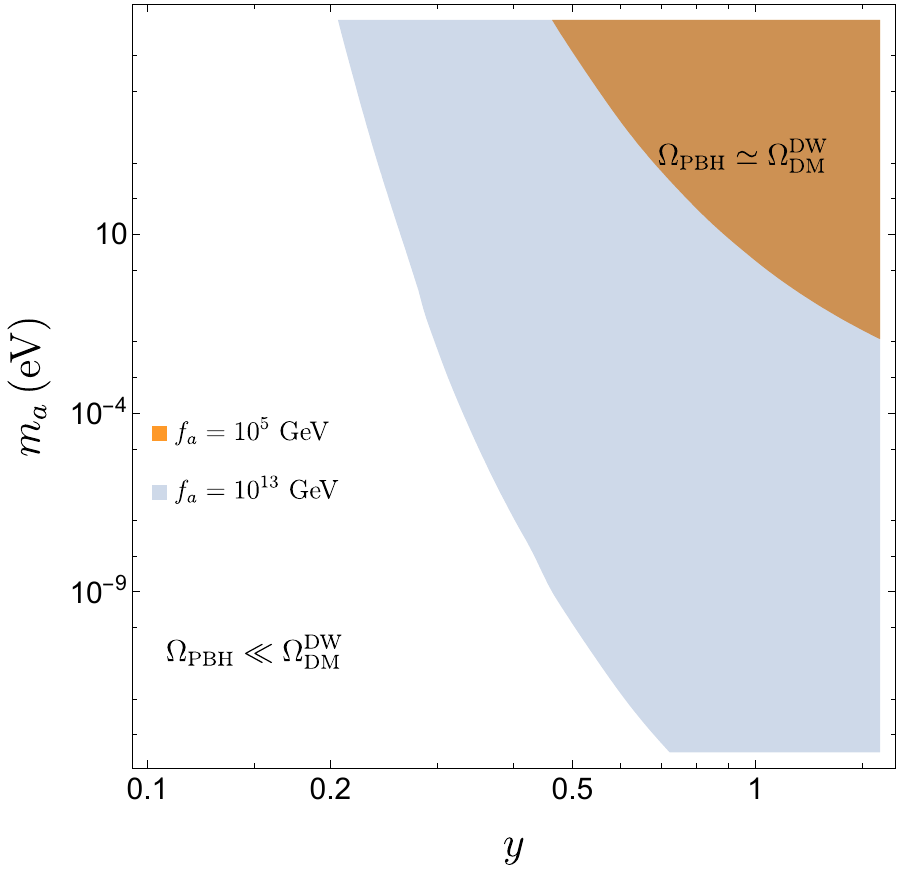}
    \caption{Comparison between two time-scales $t_\text{peak}$ vs $t_\text{PBH}$ for the QCD axion (left plot) and ALP (right plot). The colored regions represent the parameter space where $t_{\text{peak}} > t_{\text{PBH}}$ and so where most of the DWs is converted into PBHs  $\Omega_{\text{PBH}} \simeq \Omega^{\text{DW}}_{\text{DM}}$. In contrast, the non-colored region corresponds to $t_{\text{peak}} < t_{\text{PBH}}$ and the remnants of the DWs are dominated by DM axion quanta $\Omega_{\text{PBH}} \ll \Omega^{\text{DW}}_{\text{DM}}.$}
    \label{fig:tpeak vs tpbh}
\end{figure}

In Fig. \ref{fig:tpeak vs tpbh} we compare two time scales: the peak $t_\text{peak}$ of the DW distribution (the integrand in Eq. \ref{eq: Dark Matter from  Domain Walls}) and $t_\text{PBH}$ the time at which PBHs start to be formed for both the QCD axion (left plot) and an ALP (right plot). When the former is smaller, most of the DWs collapse to axion (axion-like) particles, and PBH will only give a subdominant contribution. On the contrary, when $t_\text{PBH}<t_\text{peak}$ most of the DWs enter the horizon when they are already inside their Schwarzschild radius and so most of the energy goes to PBHs. As explained in the main text, in all the regions that we identified of phenomenological interest the majority of the DM generated from the DWs was in the form of PBHs. 

\bibliography{InflationaryDWs}

\end{document}